\documentclass{aa}  

\usepackage{graphicx}
\usepackage{txfonts}
\usepackage{natbib}
\usepackage[normalem]{ulem}

\begin{document}

   \title{Pre-main sequence stars in LH 91}

   \author{R. Carini
          \inst{1}
          \and
          K. Biazzo
         \inst{1}
          \and
          G. De Marchi
  \inst{2}
          \and
          N. Panagia
  \inst{3}
          \and
          G. Beccari
  \inst{4}
         \and
         E. Brocato
 \inst{1,5}
          }

   \institute{INAF-Osservatorio Astronomico di Roma,via Frascati 33, I-00040 Monte Porzio Catone (RM), Italy\\
           \email{roberta.carini@inaf.it}
         \and
             European Space Reasearch and Technoly Centre,
            Keplerlaan 1
              2200 AG Noordwijk, Netherlands
         \and
            Space Telescope Science Institute
           3700 San Martin Drive, Baltimore MD 21218, USA
       \and
           European Southern Observatory
           Karl-Schwarschild-STr.2, 85748 Garching, Germany
      \and
          INAF-Osservatorio Astronomico d'Abruzzo,
          Via Mentore Maggini, s.n.c., 64100 Teramo, Italy
      }

   \date{}

 
  \abstract
   {}
   {We  study the accretion  properties of pre-main sequence (PMS) low-mass stars in the LH 91 association within the Large Magellanic Clouds. 
    }
   {Using optical multiband photometry obtained with the {\it Hubble Space Telescope}, we identify 75 candidates showing H$\alpha$ excess emission above  the 3$\sigma$ level with equivalent width $EW_{\rm H\alpha}$ $\geqslant$ $10$ $\AA$.  We estimate the physical parameters (effective temperature, luminosity, age, mass, accretion luminosity, and mass accretion rate) of the PMS stellar candidates.}
   {The age distribution suggests a period of active  star formation ranging  from a few million years up to $\sim$ 60 Myr with a gap between $\sim$ 5 Myr and 10 Myr. The masses of the PMS candidates span from 0.2 $M_{\odot}$ for the cooler objects to 1.0 $M_{\odot}$ with a median of $\sim$ 0.80 $M_{\odot}$.
   The median  value of the accretion luminosity of our 75 PMS stars is about  0.12 $L_\odot$, the median value of the mass accretion rate is about  4.8 $\times$ $10^{-9}$ $M_\odot yr^{-1}$ with higher values for the younger population ($\sim$ 1.2 $\times$ $10^{-8}$ $M_\odot yr^{-1}$), and lower values for the older candidates ($\sim$ 4.7 $\times$ $10^{-9}$ $M_\odot yr^{-1}$). We compare our results with findings for LH 95, the closest region to LH 91 for which accretion properties of PMS candidates were previously derived. An interesting  qualitative outcome is that  LH 91 seems  to be in a more evolved stage. Moreover, we find that the PMS candidates are distributed homogeneously, without any evidence of clumps around more massive stars.
   }

  \titlerunning{PMS stars in LH 91}
   \maketitle
%

\section{Introduction}
In the magnetospheric accretion scenario, the accretion of material from a circumstellar disk in low-mass pre-main sequence (PMS) stars is funneled by the stellar magnetic field, which disrupts the disk at  a few stellar radii (\citealt{hart}, and reference therein).
Our understanding of this process is  still not  entirely clear.
A key parameter describing the star--disk evolution is the rate of mass accretion, that is, the rate at which mass from the circumstellar disk is transferred onto the central PMS star (see, e.g.,  review by \citealt{hart}).
In particular, it is important to evaluate the relation between mass accretion rate, stellar mass,  and age, how the mass accretion rate changes as a star approaches its main sequence (MS), and how the metallicity or in general the chemical composition of the parent molecular cloud could impact the formation and evolution of the star.

Usually, mass accretion rates are derived from the analysis of continuum veiling, ultraviolet (UV) excess emission, or indeed  through a detailed study of the profile and intensity of hydrogen emission lines (e.g., H$\alpha$, Pa$\beta$, Br$\gamma$), which requires medium- to high-resolution spectroscopy for each individual object. Even with modern multi-object spectrographs at the largest ground-based telescopes, these methods  can  be applied to relatively nearby star-forming regions ($d \lesssim 1-2$ \,kpc), because of crowding. 
For this reason, the properties of low-metallicity PMS stars located in extra-galactic star-forming regions remain poorly known.

In the last decade, \cite{demarchi10} developed an efficient method based on {\it Hubble Space Telescope} (HST) photometry that allows the identification of hundreds of PMS stars simultaneously,  and the determination of their physical parameters, including effective temperature, luminosity, age, mass, H$\alpha$ luminosity, accretion luminosity,  and  mass accretion rate, with an uncertainty of between 15\% and 20\%, comparable to that allowed by spectroscopy.

This method  has been successfully applied not only to regions of the Milky Way \citep{beccari10,beccari15,zeidler16}, but also to regions of the Small \citep{demarchi11,demarchi13} and Large Magellanic Clouds (e.g., \citealt{katia19}, \citealt{demarchi17}, \citealt{spezzi12}).
This method combines $V$ (F555W) and $I$ (F814W) broadband photometry with narrow-band H$\alpha$ (F656N or F658N) imaging to identify the stars with excess in H$\alpha$  emission and to determine  their  associated H$\alpha$ emission equivalent width, $EW_{\rm H\alpha}$, the H$\alpha$ luminosity and the accretion properties of the PMS stars selected.

In this work, we use this method  to select and study the PMS populations of the stellar system LH 91 \citep{lucke74} in the  northeast outer edge of  the super-giant shell LMC\,4  in the Large Magellanic Cloud (LMC). This area, investigated with $H\alpha$ and radio observations by \cite{book},  also covers LH 91\,I in the southeast of LH 91 \citep{konti94} and LH 95 in the north of LH 91 \citep{lucke74}.

The most recent work on LH 91 was presented  by \cite{gou02} using ground-based $BVR$ and H$\alpha$ photometry. Studying the H$\alpha$ topography of the area, the authors found that LH 91 is loosely related to an \ion{H}{ii} region, which seems to be large and rather diffuse.
In agreement with \cite{lucke74}, the authors confirm that LH 91 does not seem to represent a "classical" stellar system in which  the stars are physically related to each other.
Analyzing the color--magnitude diagram (CMD) in the $B$ and $V$ band, the authors estimated the color excess  $E(B-V)$ = 0.16 $\pm$ 0.04 using  the reddening-free Wesenheit function. Moreover, fitting the Geneva isochrones \citep{geneva} derived adopting metallicity Z=0.008, \cite{gou02} derived the age of the system, finding it to be younger than 10 Myr, similar to that of LH 95 and LH 91I, and in agreement with for example \cite{braun97,braun00}.
Instead, \cite{konti94} estimated an age of  about 20 Myr.
Finally, \cite{gou02} also estimated the age of the background field, the  population of the observed area around LH 91, to be older than 50 Myr and up to 1.25 Gyr.

This paper is organized as follows:
in Section \ref{phot} we describe the HST photometric observations, in Section \ref{idpms} we illustrate the analysis needed to identify the PMS stars and to estimate the luminosity  and the equivalent width (EW) associated to the H$\alpha$ excess. In Section \ref{physpam} we measure the physical properties of the stars selected. In Section \ref{proacc} we determine the accretion properties of the selected PMS stars,  that is, the accretion luminosity and mass accretion rate, and we show the relation between the mass accretion rate and the stellar properties of the PMS objects, such as their mass and age. We also compare our results with the findings for other star-forming regions in the LMC with the same metallicity, and in particular with LH 95, the closest region to LH 91 for which accretion properties of PMS candidates have been derived \citep{katia19}. We present our conclusions in the last section.

\section{Photometric observations} 
\label{phot}

The LH 91 region was observed with  the Wide Field Camera 3 (WFC3/UV) on board the HST in the broad-band filters $F555W$ and $F814W$, and in the narrow-band filter $F656N$, the latest centered on the H$\alpha$ line. 
The data were collected as part of HST programs \#12872 (PI: Da Rio) and
\#13009 (PI: De Marchi). A short logbook of the observations is shown in Table~\ref{tab_obs}.

\begin{table}

 \centering
  \caption{Logbook of the observations}
  \begin{tabular}{@{}cccc@{}}
  \hline
  Camera    & Number of exposures & Filter & Exposure time (s) \\
            &          &       &  (s) \\
 \hline
\multicolumn{4}{c}{Prop ID 12872, PI: Da Rio}\\
 \hline
  WFC3 & 2 &  $F555W$ & 2804 + 2970 \\
       & 1 &  $F814W$ & 2804 \\
       & 1 &  $F656N$ & 2970 \\
 \hline
\multicolumn{4}{c}{  Prop ID 13009, PI: De Marchi}\\
 \hline
  WFC3 &  1 & $F656N$ & 2949 \\
  \hline
\label{tab_obs}
\end{tabular}

\end{table}

The data were reduced using the standard $DAOPHOTII$ \citep{stetson} procedure. A list of 10 to 20 well-sampled and isolated stars were used to model the point spread function (PSF) on the $F555W$ and $F814W$ images, and a deep photometric catalog of stars was derived via PSF fitting on the images acquired with the broad-band filters. The final magnitude of each star in a given filter is estimated as the mean of the photometric measures in each individual image taken with that filter, while the standard deviation is taken as the associated error. Aperture photometry was then used
to extract the $F656N$ magnitude for each star detected in the optical bands.
The choice of performing aperture photometry on the narrow-band images is driven by the fact that such images are characterized by very little stellar
crowding. As such, the aperture photometry is the ideal choice as it allows accurate estimation of the magnitude free from any uncertainty that is unavoidably associated with the choice of PSF model. We stress here that the background is locally estimated and subtracted in an annulus around the start.
The final catalog of the overlapping fields contains 9423 objects,  of which  6980 have a measure in the $F656N$ band.
These $F555W$ and $F814W$ band observations are among the deepest ever taken toward the LH 91 region.
The instrumental magnitudes in $F555W$, $F814W,$ and $F656N$ were calibrated to the
VEGAMAG photometric system using the zero-point values made available by the Space Telescope Science Institute\footnote{https://www.stsci.edu/hst/instrumentation/wfc3/data-analysis/photometric-calibration}.\\

\section{Data analysis}
\label{idpms}

\subsection{PMS star identification}
\label{analisi}
We applied the method developed by \cite{demarchi10} to identify the PMS stars characterized  by an active mass accretion process. We  measured the physical and accretion properties of these objects (i.e., H$\alpha$ luminosity, H$\alpha$ emission $EW_{\rm H\alpha}$, mass accretion rate, and accretion luminosity) using photometric data. We refer to \cite{demarchi10} for a detailed discussion of the method, while in this work we describe some fundamental steps.

\begin{figure}
   \includegraphics[width=10cm]{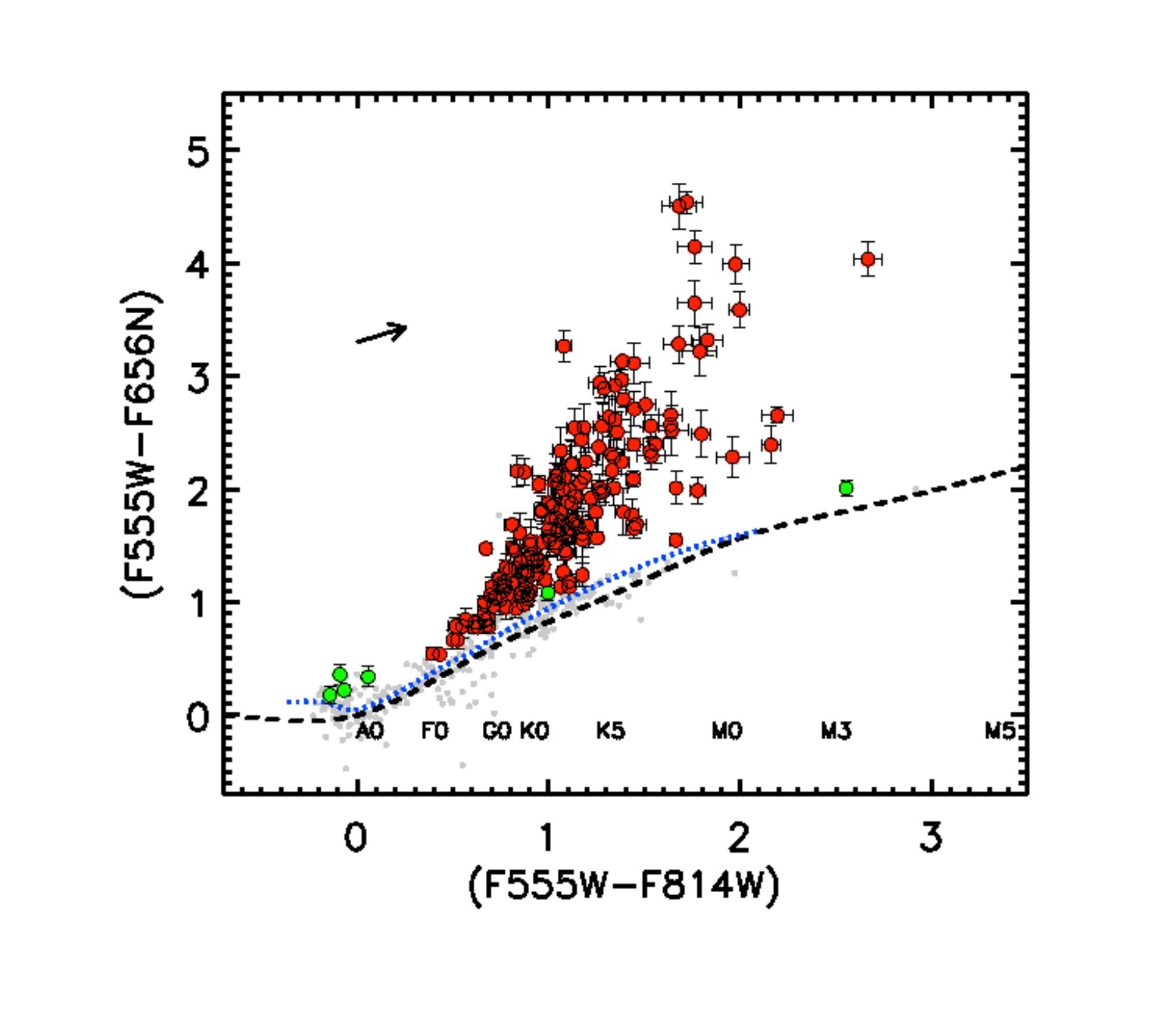}
      \caption{Color--color diagram of the selected  stars in the field of LH 91. All magnitudes are already corrected for  the extinction contribution of our Galaxy, $A^{MW}_{555}$ = 0.22 mag and $E(m_{555}-m_{814})^{MW}$ = 0.1.  The arrow shows the  reddening vector of $E(m_{555}-m_{814})$ =0.25 and $E(m_{555}-m_{656})$=0.13 for the adopted LH 91 extinction law.
     The dashed line represents the  median photospheric ($m_{555}-m_{814})$ color for the 254 stars with $\delta_{555}$, $\delta_{814}$, and $\delta_{656}$ < 0.05 (gray filled dots).
   The dotted line shows the model atmospheres of \cite{bessel}   computed for the three WFC3/UVES filters.
The PMS star candidates with H$\alpha$ emission excess at the $3\sigma$ level are represented with large red dots. 
The large green dots are the brightest PMS star candidates, with $m_{555}$ < 20 mag. Error bars are also shown. 
 }
         \label{Vi}
   \end{figure}

We selected PMS stars on the basis of their H$\alpha$ excess emission \citep{white03}.
First of all, we identified the H$\alpha$ excess emitters  in the $(m_{555}-m_{656})$ versus  
$(m_{555} - m_{814})$ color--color diagram  shown in Fig. \ref{Vi}. The magnitudes were corrected for the extinction contribution of the Milky Way considering the values $A^{MW}_{555}$ = 0.22 mag and $E(m_{555}-m_{814})^{MW}$ = 0.1 \citep{fitz}.
To this aim, we selected from our catalog in $F555W$,  $F814W$, and $F656N$ bands all those stars whose photometric uncertainties, that is $\delta_{555}$, $\delta_{814}$, and $\delta_{656}$,  are less than 0.05 in each individual band.
A total of 254 stars satisfied these conditions (gray filled dots in the color--color diagram in Fig.\ref{Vi}), out of 9423 sources in the whole catalog. These are typically MS stars that do not present an appreciable H$\alpha$ excess.
With these stars, we define a reference sequence (dashed black line) with respect to which the excess H$\alpha$ emission is computed.
The dotted blue line  of Fig. \ref{Vi} represents  the theoretical color relationship obtained using the \cite{bessel} model atmospheres for MS stars with the chemical and physical parameters appropriate for the LMC (effective temperature  $T_{\rm eff}$ in the range of 3500-40000 K, surface gravity $\log g$ = 4.5, and metallicity index $[M/H]$ $\simeq $ -0.5, \citealt{colucci}). The  agreement between our reference sequence and the theoretical one is evident at $m_{F555W}-m_{F814W}$ < 1.
The discrepancy between the models and the data at $m_{F555W}-m_{F814W}$ > 1 can be attributed to small number statistics and to  the fact that the majority of these objects are red giants, with different physical characteristics  from those assumed in the models.

To select the most probable PMS stars, after the exclusion of the 254 stars taken as reference, we first selected the targets with photometric uncertainties in each individual band as follows:  $\delta_{555}$ and $\delta_{814}$  $<$ 0.1 mag, and $\delta_{656} <$  0.3 mag, for a total of 1309 objects.
As highlighted by \cite{demarchi10}, the contribution of the H$\alpha$ line to the $m_{555}$  magnitude is negligible, and therefore we can define the magnitude of the excess emission as:
\begin{equation}
\Delta H\alpha= (m_{555}-m_{656})^{obs}-(m_{555}-m_{656})^{ref}
,\end{equation}
where the superscripts "obs" and "ref" refer to the observation and reference sequence, respectively.
We then considered the stars   with $\Delta H\alpha$ exceeding  at least three times  the  combined mean  photometric uncertainties in the three bands  $\delta_3$:

\begin{equation}
\delta_3=\sqrt{\frac{\delta_{555}^{2}+\delta_{656}^2+\delta_{814}^2}{3}}
.\end{equation}

A total of 187 stars satisfy these conditions; they are indicated  with large red dots in Fig \ref{Vi}.
This means that 187 stars have $(m_{555}-m_{656})$ colors exceeding that of the reference template at the given $(m_{555}-m_{814})$ color by more than three times the combined uncertainties on their $(m_{555}-m_{656})$ values.
The large green dots in Fig.1 are the targets selected with $m_{555}$
< 20 mag, which we exclude  from our following analysis as we are interested primarily in low-mass PMS candidates. Our final sample of PMS candidates is therefore composed by 181 targets.
As in these bands the reddening vector due to LH 91 runs almost parallel to the median of the reference sequence \citep{demarchi10}, the color--color diagram  provides  a robust identification of stars with H$\alpha$ excess even before correction for LH 91 reddening.

%

\subsection{The color--magnitude diagram}
\label{reddening}

\begin{figure}
   \centering
   \includegraphics[width=10cm]{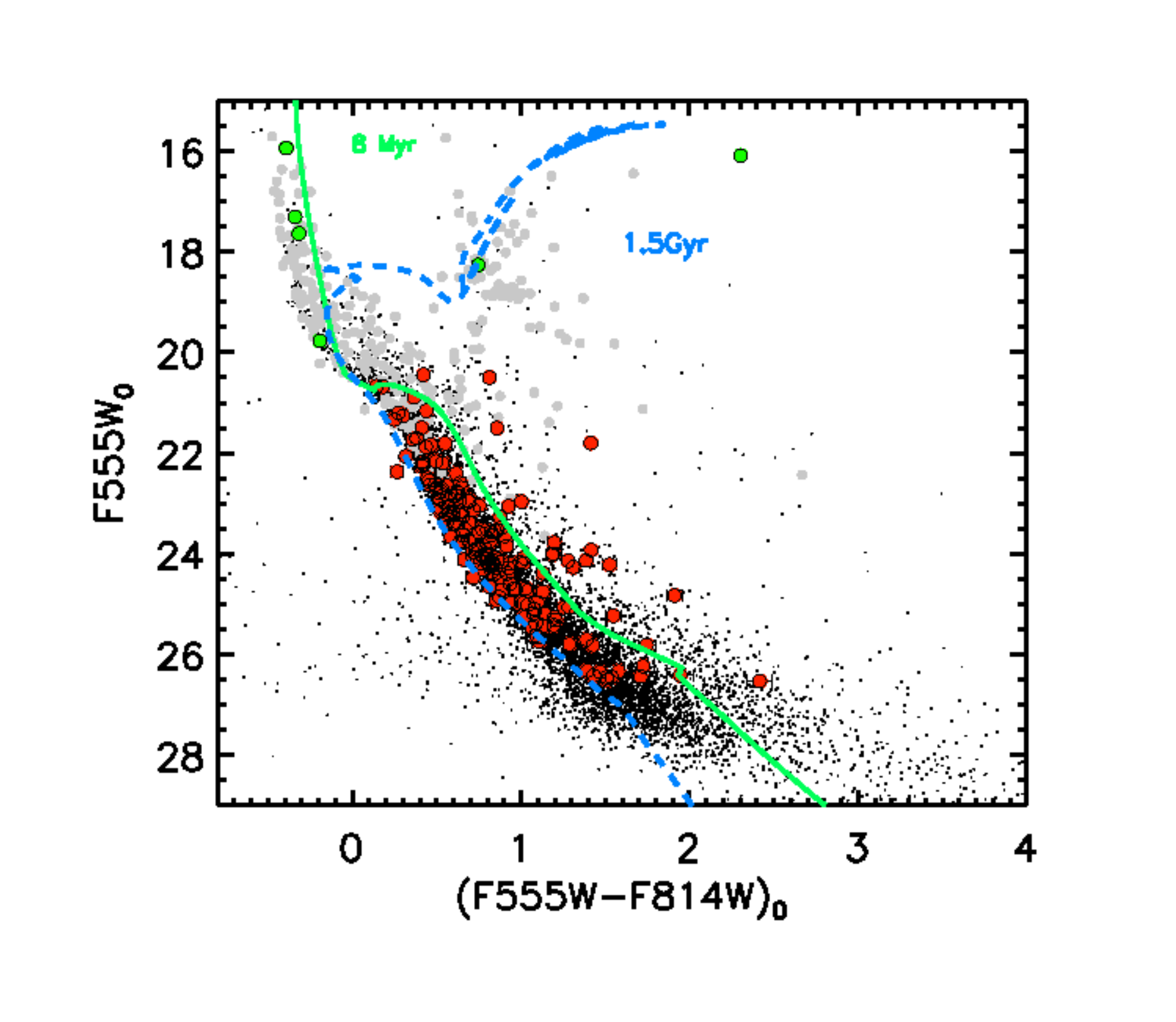}
      \caption{Color--magnitude diagram of the field of LH 91. All magnitudes are already corrected for  the extinction contribution of our Galaxy and LH 91.  The small black dots are the  targets of the whole sample (9423 stars).
Small gray-filled dots  are the stars with photometric uncertainties < 0.05 mag in each band.   The large red dots represent the PMS star candidates with H$\alpha$ excess emission at the $3\sigma$ level.
       Solid green and dashed light-blue  lines  show the theoretical isochrones from  \cite{bressan2012} for ages 8 Myr  and 1.5 Gyr, respectively, metallicity Z=0.007, and a distance modulus $(m_V -M_V)_0$=18.55.}
   \label{cmd}
   \end{figure}

We applied the correction for the extinction contribution of the Milky Way and LH 91 to the magnitudes in each band.
For the Milky Way,  we report   the values in the previous section. 
We estimated the extinction for LH 91 from the value of $E(B-V)$ = $0.16$ $\pm$ $0.04$  color excess in the photometry of \cite{gou02} and converted into $A_V$ assuming the average LMC reddening law  $R_{555}$ = $A_{555}/E(m_{555}-m_{814})$ $=$ 2.97 calculated by \cite{demarchi14}. 
As \cite{gou02} found that the density of the ambient medium in LH 91 is similar to the value  for LH 95,  and as \cite{dario09} did not find a significant  level of differential extinction while studying the upper
MS stars of the latter,  we also consider the differential reddening to be negligible in LH 91.
We show the CMD $(m_{555}-m_{814})_0 $ versus $(m_{555})_0$ in Fig. \ref{cmd}. The small black dots are the  targets of the whole sample, namely 9423 stars.
To estimate the age of the system, we fit the CMD with the isochrone models for Z=0.007 ---which is typical of young LMC stars (e.g., \citealt{colucci})--- taken from the PAdova-Trieste Stellar Evolution Code (PARSEC, \citealt{bressan2012}) and distance modulus $(m-M)_0$=18.55 \citep{panagia91,panagia99}.
The turnoff at $m_{555} \sim 20.5$\,mag and the red clump  at $m_{555} \sim 19.5$\,mag and $m_{555}-m_{814} \sim 1.0$ are best matched by a 1.5 Gyr isochrone (dashed light-blue  line), in agreement with the age of the background field stars evaluated by \cite{gou02}.
Stars with  H$\alpha$ excess show a wide apparent spread towards young age and could be  divided  in two groups, separated by an isochrone at 8 Myr (solid green line). 

\subsection{From H$\alpha$ color excess to H$\alpha$ luminosity}
\label{Ha_luminosity}

To avoid contamination by stars with significant chromospheric activity, we also imposed constraints on $EW_{\rm H\alpha}$, selecting only stars with  $EW_{\rm H\alpha}$ $\geqslant$ 10 \AA, because according to  \cite{demarchi10} this  is a reliable cutoff to separate accretors from those not accreting .

For details of the method used here to derive $EW_{\rm H\alpha}$ from the photometry,  we refer to \cite{demarchi10,demarchi11,demarchi13}. Here we recall that, as the  width of the H$\alpha$ line is narrow with respect to the width of the filter, the measure of $EW_{\rm H\alpha}$ is given by the difference between the observed H$\alpha$ line magnitude and the level of the H$\alpha$ continuum ($\Delta H\alpha$).
If we assume that the stars used to define the reference sequence have no H$\alpha$ absorption features, their $(m_{555} - m_{656})$  color represents  the color of   the pure continuum.
Consequently,  we calculated the $EW_{\rm H\alpha}$ from the following relation:
\begin{equation}
EW_{\rm H\alpha}=RECTW \times [1-10^{-0.4\Delta H\alpha}]
,\end{equation}

\noindent{where $RECTW$ is the rectangular width of the F656N filter. The uncertainty on the $EW_{\rm H\alpha}$ measure is dominated by the uncertainty on the H$\alpha$ magnitude.}

\begin{figure}
   \centering
   \includegraphics[width=10cm]{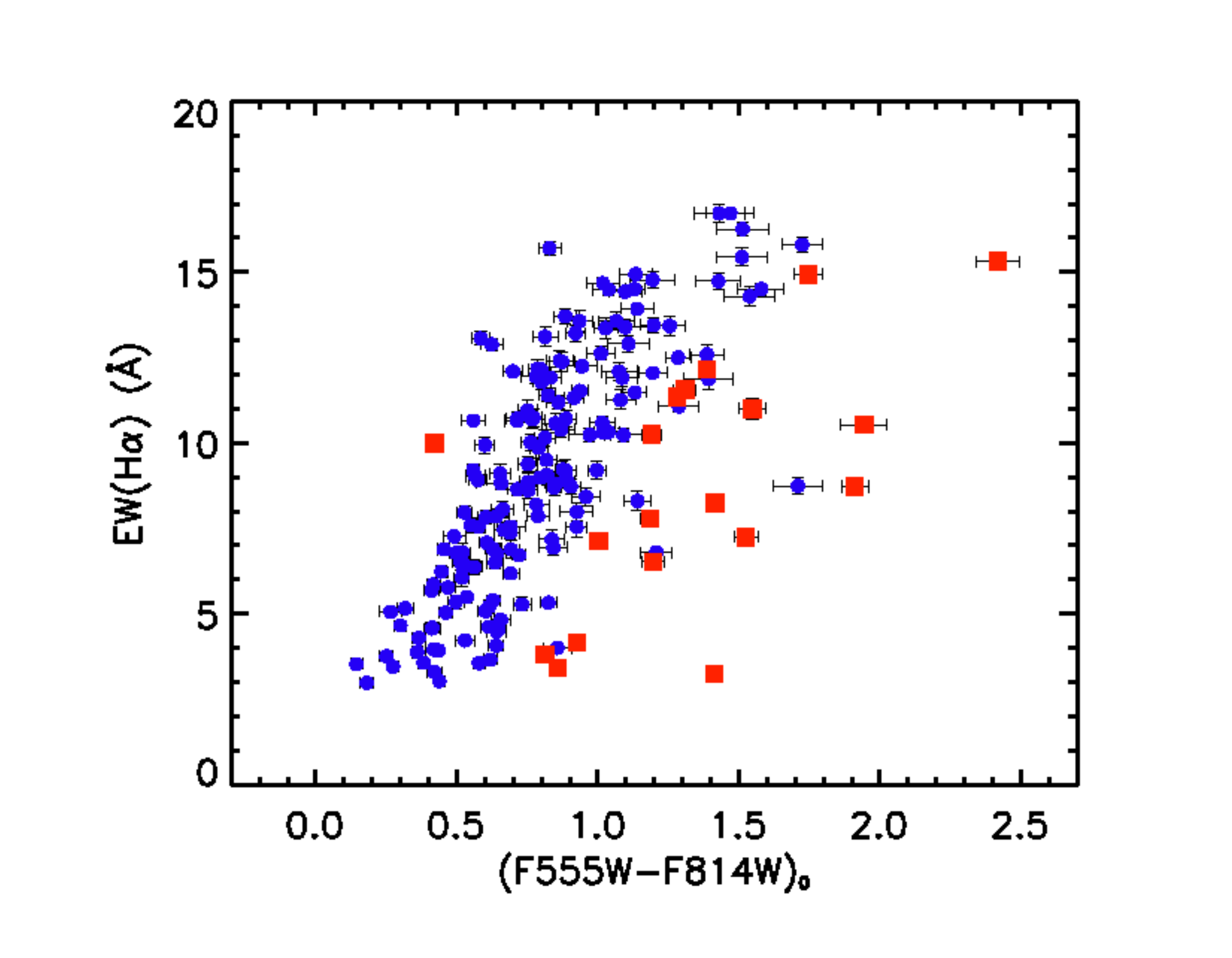}
      \caption{H${\alpha}$ equivalent width of the selected low-mass PMS candidates, as a function of the de-reddened ($m_{555}-m_{814}$) color. The red squares represent the values of the PMS stars younger than 8 Myr, the blue dots are the oldest PMS stars.
      } 
   \label{ew}
   \end{figure}

Moreover, because of the width of the F656N filter,  the small contribution due to the emission of the forbidden $\ion{N}{ii}$ line at $\lambda 6548$  is included in $\Delta H\alpha$. 
Therefore, following the prescriptions by \cite{demarchi10}, we estimated corrections ranging from 0.2  to 1.4 $\AA$, to be subtracted from the $EW_{\rm H\alpha}$ of our targets.
Figure \ref{ew}  shows the $EW_{\rm H\alpha}$  measured for the selected low-mass PMS candidates as a function of the de-reddened  $m_{555}-m_{814}$ color. 

We performed a preliminary study of the $EW_{\rm H\alpha}$ distribution  of  the PMS  candidates at different ages using the isochrone at 8 Myr as a discriminating factor. We divided the sample into stars  older (blue dots)  and  younger (red squares)  than 8 Myr (Fig. \ref{ew}). 

The values of $EW_{\rm H\alpha}$ for the sample range from $\sim$ 3 $\AA$ to $\sim$ 17 $\AA$, with a  median of $\sim$ 9 $\AA$ that applies to both the whole sample and the two subgroups.
After the selection on the $EW_{\rm H\alpha}$, a total of 75 objects  satisfy the conservative condition ($EW_{\rm H\alpha}$ $\geq$ 10 $\AA$). The median value of the $EW_{\rm H\alpha}$ is about 12 $\AA$, regardless of age, smaller than the values found in other star formation fields in the LMC, such as LH 95 ($EW_{\rm H\alpha}$ $\sim$ $30$ $\AA$, \citealt{katia19}) and SN 1987A ($EW_{\rm H\alpha}$ $\sim$ $20$ $\AA$, \citealt{demarchi10}).
The difference could be due to the paucity and to stellar mass range of our sample.
Moreover, the figure shows an almost clear separation in color between the two subgroups in LH 91 with the exception of the target with $(m_{555} - m_{814})_{0}$ $\sim$ $0.4$ and $EW_{\rm H\alpha}$ $\sim$ $10$ $\AA$. As the coordinates of this target correspond to those of a massive star in the 2MASS catalog, it could be a Be star. A similar separation in color between the two subgroups of PMS stars was found in LH 95 \citep{katia19}.

The H$\alpha$ emission line luminosity $L_{\rm H\alpha}$ can be determined  from the absolute sensitivity of the instrumental setup, the photometric zero point (ZP), the distance of the stars, and from the magnitude in the H$\alpha$ band:

\begin{equation}
L_{\rm H\alpha}=4\pi d^210^{0.4(ZP-m_{656})}\rm{PHOTFLAM}\times \rm{RECTW} 
.\end{equation}
The  values of the photometric properties of the instruments were taken from  \cite{ryon18}, namely  the inverse sensitivity PHOTFLAM= 1.714 $\times$ $10^{-17}$ erg cm$^{-2}$ s$^{-1}$ \AA$^{-1}$,  and the zero point in the VEGAmag system for the H$\alpha$ filter, ZP= 19.84 \citep{calamida}.
Assuming a distance of 51.4 $\pm$ 1.2 kpc \citep{panagia99} and considering the rectangular width of the  $F656N$  filter RECTW= 17.679 \AA, we determined the  H$\alpha$ luminosity for the 75 targets, finding a  median value of about 8.7 $\times$ $10^{30}$ erg s$^{-1}$ ($0.2 \times 10^{-2} L_\sun$).

This value is slightly lower than the one found by \cite{katia19} in the LH 95 association ($\sim$ 1.2$\times$ $10^{31}$ erg s$^{-1}$, $ 0.3 \times 10^{-2} L_\sun$).
This  is not surprising  because
\cite{gou02} found that the mean $H\alpha$ intensities of  the HII region related to LH 95 (DEM L 252) is about two times higher than the corresponding intensity of the HII region associated with LH 91 (DEM L 251). 
 \\
 In addition, we can compared our result with the median H$\alpha$ luminosity of other regions of the LMC, namely 30 Doradus Nebula
 and SN 1987A field; also in these cases our value is lower,  the mean L$_{\rm H\alpha}$ estimated in these regions is $\sim$ 4 $\times$ $10^{31}$ erg s$^{-1}$($\sim$ $10^{-2} L_\sun$, \cite{demarchi10}) and  $\sim$ 1.5$\times$ $10^{32}$ erg s$^{-1}$ ($\sim$ 4 $\times$ $10^{-2}$ $L_\sun$ \cite{demarchi17}) respectively.


The uncertainty on $L_{\rm H\alpha}$ is dominated by the uncertainties on the H$\alpha$ photometry, on the distance ($\sim$ 5\%) and on instrumental setup ($\sim$ 3\%) (see \citealt{demarchi10}).
The total uncertainty on $L_{\rm H\alpha}$ is about 16 \%.

\section{Physical parameters of the PMS candidates}
\label{physpam}
\subsection{Effective temperature and bolometric luminosity}
We  evaluated the effective temperature of the PMS candidates by comparing the theoretical models  with the $m_{555}-m_{814}$  color  of our sample corrected for the reddening due to the Milky Way and LH 91, as explained in Sect. \ref{reddening}.
To convert the color to $T_{\rm eff}$ we used the models of \cite{bessel} for  3500 K $\leq$ $T_{\rm eff}$ $\leq$ 40000 K, $\log g$=4.5, and metallicity index [M/H] =-0.5 dex.
As the models of \cite{bessel}  are not available for temperatures lower than 3500 K, we used the $T_{\rm eff}$-$(V-I_C)$ calibration by \cite{mamajek}, with the assumption that the calibrated $m_{555}$ and $m_{814}$ magnitudes coincide with the $V$ and $I_C$ magnitudes (see \citealt{katia19}).

\begin{figure}
   \centering
   \includegraphics[width=10cm]{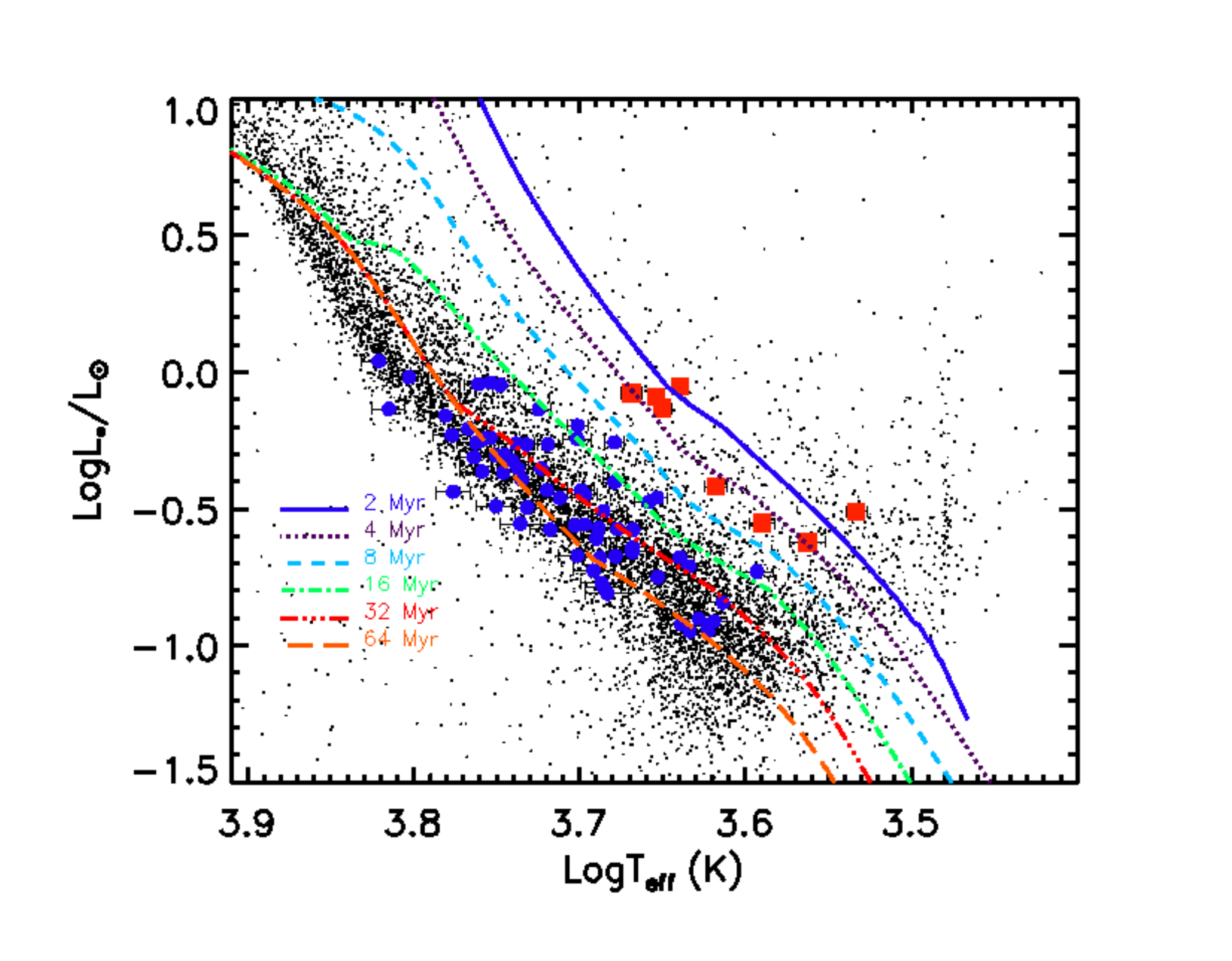}
      \caption{HR diagram with the location of the low-mass  young (red squares) and old (blue dots) PMS candidates. The  theoretical  isochrones \citep{bressan2012} are  calculated for ages of 2, 4, 8, 16, 32, and 64 Myr (lines from right to left, respectively) and Z=0.007. The small black dots are the targets of the  sample of our catalog. 
      } 
   \label{hr}
 \end{figure}

To obtain the luminosity of the stars $L_{\star}$, we considered the magnitude $m_{555}$ corrected for the interstellar extinction, a distance  to LH 91  of  51.4 kpc \citep{panagia99}, and a bolometric solar magnitude of 4.74 mag \citep{mamajek}.
The uncertainty on the  effective temperature  and stellar luminosity  are dominated by the uncertainties on the magnitudes and distance.
In Fig. \ref{hr},  we show the location of the PMS candidates in the HR diagram, with the relative uncertainties, which in some cases  are smaller than the symbol size.
We highlight that the majority of the PMS candidates are close to the MS and we could only identify them thanks to the information on their  H$\alpha$ excess.
We also plot in Figure 4 the theoretical  isochrones for ages of  2, 4, 8, 16, 32, and 64 Myr  for Z=0.007 \citep{bressan2012}.
The red squares represent the PMS candidates younger than 8 Myr, while the blue dots the older ones.
 
From the HR diagram, it appears that LH 91 is characterized by a more or less continuous star formation, from a few million years to $\sim$ 60 Myr,  with a smaller number of PMS candidates for ages younger than 8 Myr.

From the effective temperature and the luminosity of the PMS stars, we derive the stellar radius  $R_\star$ of these stars, which we use to estimate the mass accretion rate of the selected PMS objects in Section\,\ref{proacc}. Typical mean errors on $R_\star$ are around 7\% and include both uncertainties on $T_{\rm eff}$ and $L_\star$.

\subsection{Mass and age}
\label{mage}

\begin{figure}
   \centering
   \includegraphics[width=10cm]{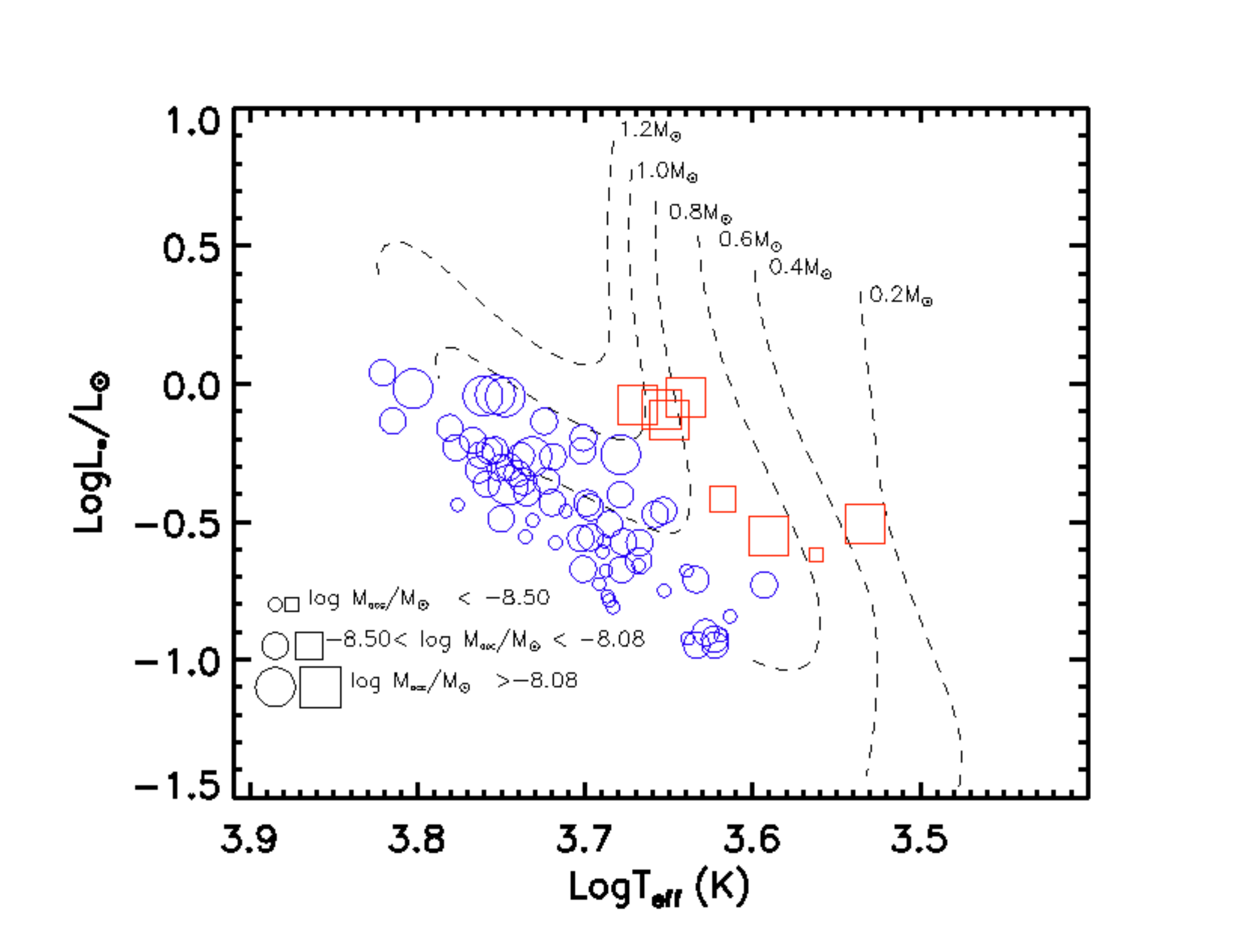}
      \caption{HR diagram of our PMS candidates. Red squares and blue dots represent young and old PMS stars, respectively. The size of the symbols is proportionate to the rate of mass accretion, as in the legend. We adopted the PARSEC tracks for metallicity $Z=0.007$  \citep{bressan2012} from 0.2 $M_\sun$ to $1.2$ $M_\sun$ (dashed lines).
      } 
   \label{tracce}
\end{figure}

We derived the  mass and age for each target by comparing the location of each star in the HR diagram (Fig. \ref{tracce}) with theoretical PMS evolutionary tracks. We adopted the PARSEC tracks for metallicity $Z=0.007$ \citep{bressan2012} from 0.1 $M_\sun$ to $3.0$ $M_\sun$.
We followed the approach discussed in \cite{Romaniello98} and refined by \citet{demarchi11,demarchi13}.
According to these authors, we define a grid in luminosity and temperature consisting of evenly spaced cells with sizes comparable to the typical observational errors. Given an evolutionary track of a star of a certain mass, we identify the cells crossed  by the star during its evolution. For each cell, we extrapolate information associated with the evolutionary track, namely mass and age. The information is then be associated with the observed star belonging to a particular cell (for further details, see \cite{demarchi17}).
Figure \ref{tracce} shows the masses  of the PMS candidates spanning from $0.2\,M_\odot$ for the cooler objects up to $1.0\,M_\odot$ for the hottest ones. The median value of the sample is about $0.8 M_\odot$. In the figure, we divided the PMS stars in two subsamples: the younger PMS candidates with an age of less than 8 Myr (red squares), and the PMS candidates older than 8 Myr (blue dots).
The sizes of the dots  and squares are proportional to  the mass accretion rate, which we determine and discuss in Section\,\ref{proacc}.
Here, we simply want to investigate whether and  how  the rate of  mass accretion is correlated with evolutionary phase and stellar mass. As one can see in Fig.\,\ref{tracce}, the targets with the highest mass accretion rates (the largest symbols) are the youngest PMS stars, while the mass accretion rate decreases  at older ages. Furthermore, the stars with higher mass  have higher mass accretion values ($\dot{M}_{\rm acc}$) at all ages.

\begin{figure}
   \centering
   \includegraphics[width=10cm]{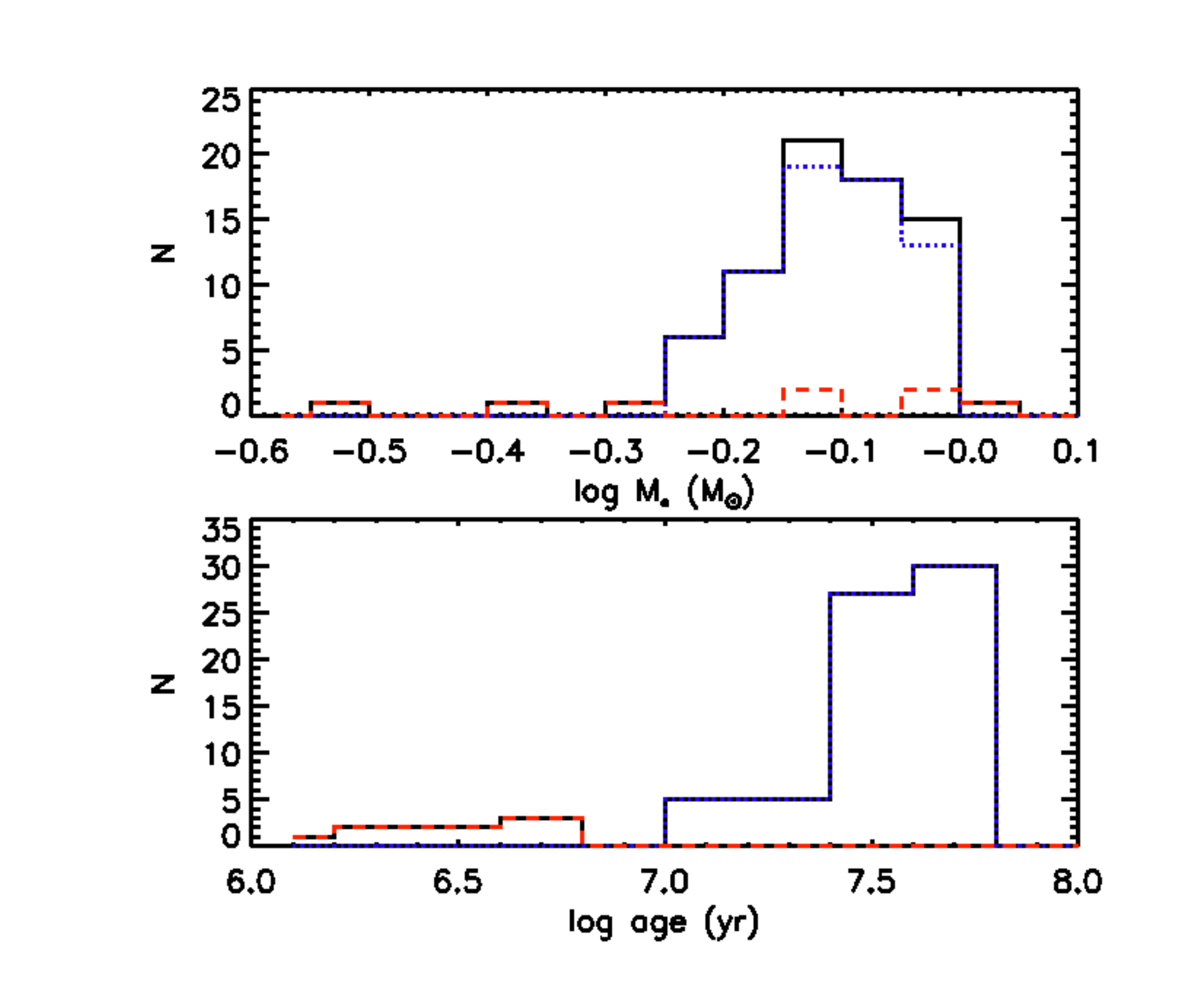}
      \caption{Histograms of the stellar mass (upper panel, bin of 0.05) and age (lower panel, bin of 0.2) for the 75 low-mass PMS candidates in logarithmic scale. The red dashed and blue dotted  lines represent the distribution of the young and older populations, respectively.
      } 
   \label{histo}
\end{figure}

In Fig. \ref{histo} we show the histograms of the mass (upper panel) and age (lower panel) distribution of the PMS candidates with the bin sizes compatible with the uncertainties on mass and age, respectively. 
The black line corresponds to the whole sample,  the dashed red line corresponds to the young PMS candidates, while the dotted blue line represents the old PMS candidates.
The older PMS stars include preferentially higher mass stars, with the mass distribution presenting a peak at $\sim$ 0.7 $M_\odot$.  The young PMS objects show  a continuous distribution in mass, with no  evident   peak, but this is probably mostly due to the paucity of the subsample.

The age distribution could suggest a separation between older and younger PMS stars, with a gap in the range between 5 and 10 Myr. The younger population shows a continuous distribution in age up 5 Myr. The older population constitutes about 90\% of the objects, with ages between 10 and $\sim$ 60 Myr and a peak at $\sim 50$\,Myr.

\section{Accretion properties}
\label{proacc}
In the following subsections we  describe how we determined the accretion properties of our sample of PMS candidates and we present our study of their relation with the  physical properties of the stars.
\subsection{Accretion luminosity}
\label{acclum}
The luminosity of the H$\alpha$ line generated along the funnel flows of  circumstellar gas during the magnetospheric accretion process can be used as a tracer to  estimate the accretion luminosity.
To determine the accretion luminosity of our sample of PMS candidates, we adopted the relationship  obtained by \cite{demarchi10}, who  analyzed the data of a group of T Tauri stars in Taurus-Auriga  compiled by \cite{dahm}:

\begin{equation}
\log {\frac{L_{\rm acc}}{L_\odot}}= \log{\frac{L_{\rm H\alpha}}{L_\odot}} + (1.72 \pm 0.25)
.\end{equation}

The median of the accretion luminosity of our 75 PMS stars  is 0.12 $L_\odot$.
The uncertainty on $L_{\rm acc}$ is dominated by the uncertainty on $L_{\rm H\alpha}$, which is about $16\%$, related to the photometric error on the H$\alpha$ magnitude. There is also  a systematic error to take into account due to the  uncertainties on the ratio  $L_{\rm acc}/L_{\rm H\alpha}$ \citep{dahm,demarchi11}, but  as the relation is the same for all stars,  this uncertainty  does not interfere with the comparison between the targets.

\begin{figure}
   \centering
   \includegraphics[width=10cm]{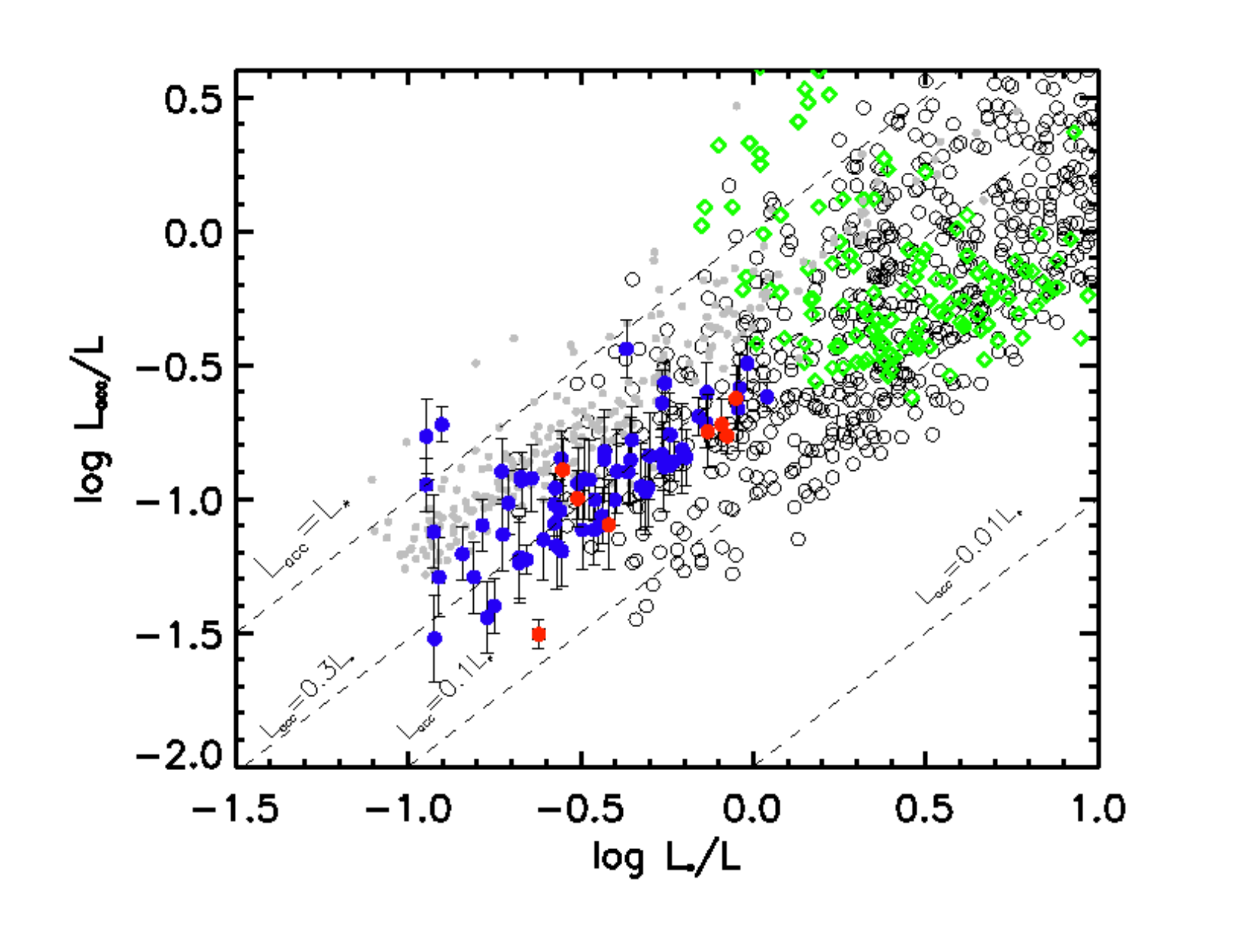}
      \caption{Accretion luminosity as a function of  stellar luminosity. Blue  and red dots represent the older (age greater than 8\,Myr)   and younger (age smaller than 8\,Myr)  PMS candidates of LH 91, respectively. The gray filled dots, green diamonds, and black empty dots are  the PMS of LH 95 by \cite{katia19},  SN 1987A by \cite{demarchi10}, and 30 Dor by \cite{demarchi17}, respectively. The dashed lines show the linear $L_{\rm acc}-L_\star$ relationship for different values of the coefficient, as indicated.   } 
   \label{lacc}
   \end{figure}

In Fig. \ref{lacc} we show the accretion luminosity versus $L_\star$ of the PMS candidates, the blue dots and red squares representing  the old  and young  ones, respectively.
In each star formation region, $L_{\rm acc}$ increases with stellar luminosity, but the range and dispersion of the data are quite different.
For comparison, we show also the data of LH 95, with gray filled dots, SN 1987A with green empty diamonds, and 30 Dor with black empty dots.
In LH 91 and LH 95, the dispersion in $L_{\rm acc}$ seems to decrease with the increase of $L_\star$. 
The accretion luminosity spans a range between 0.1 and 1 $L_\star$, with the peak of the distribution at about $0.3\,L_\star$ for LH91. In Section \ref{Ha_luminosity}, we shown that the median $L_{\rm H\alpha}$ in LH 91 is lower than that found in LH 95, and therefore it is not surprising that the values of the accretion luminosity in LH 91 are 
also slightly lower than those of the PMS objects in LH 95 ($\sim$ $0.17$ $L_\odot$). This result could be due to two main factors: in LH 95 the mass range of the sample is larger (0.2-1.8 $M_\odot$), and at the same mass the stars are younger.
 The samples  of 30 Dor and SN 1987A are  richer than LH 91 and LH 95, and the range of the  accretion luminosity is  larger, from  0.1 $L_\star$ to values higher than 1.0 $L_\star$.  For a comparison, we focus on the range in stellar luminosity in common between the regions, namely  -0.65 $L_\star$ and 0.0 $L_\star$. We evaluated the median accretion luminosity only for the regions 30 Dor, LH 95, and LH 91  finding values of $\sim$ 0.17 $L_\odot$, 0.22 $L_\odot$, and 0.13 $L_\odot$, respectively.
Unfortunately, the range in star luminosity of SN 1987A does not cross-match  with those of LH 91 and LH 95, and therefore we cannot make a
direct comparison.

\begin{figure}
   \centering
   \includegraphics[width=10cm]{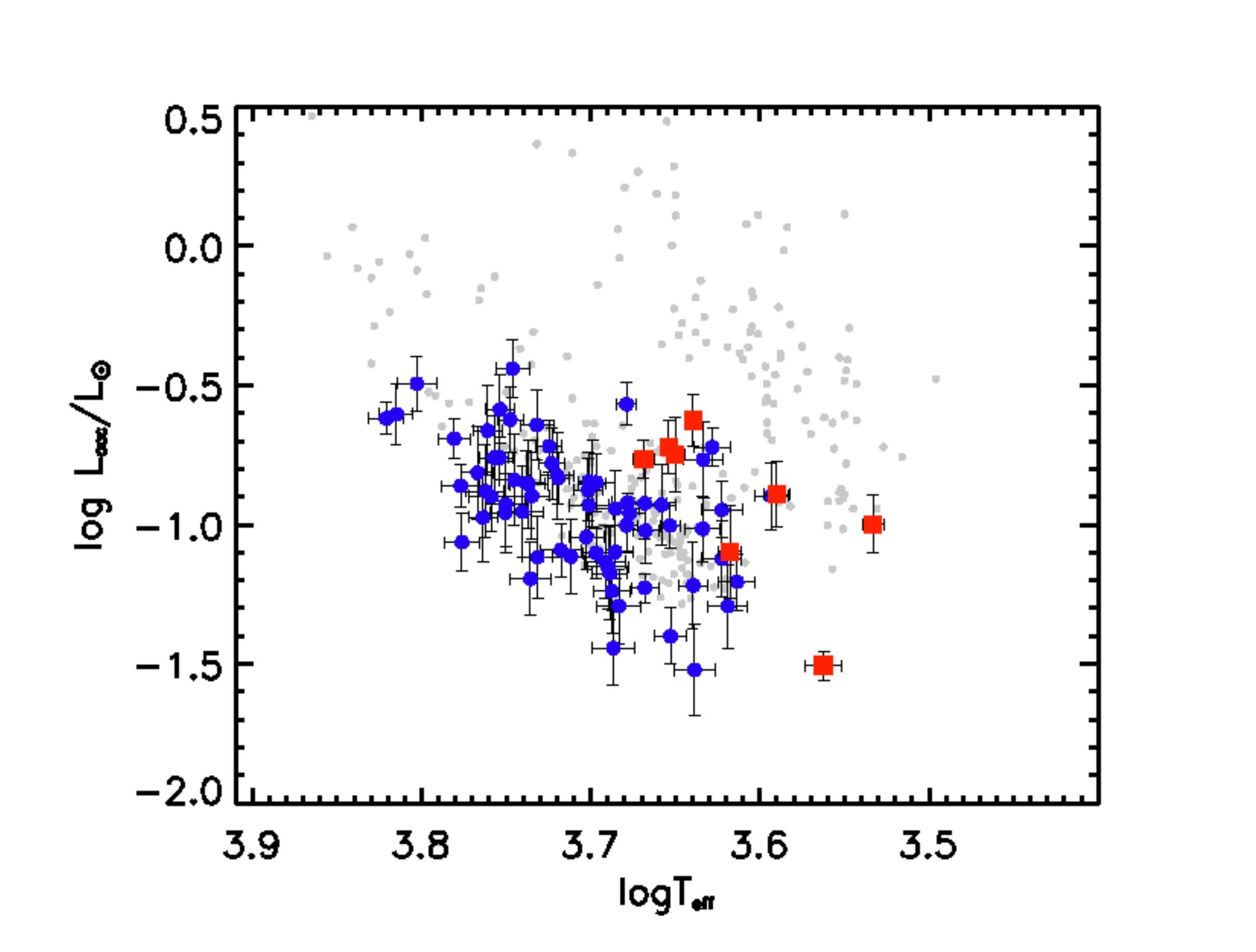}
      \caption{Accretion luminosity versus effective temperature. The blue dots and red squares  are as in Fig. \ref{lacc}, the gray filled dots  represent the PMS stars of LH 95 by \cite{katia19}. 
      } 
   \label{tef}
   \end{figure}

In Fig. \ref{tef} we show the accretion luminosity versus the effective temperature  in logarithmic scale of the old (blue dots) and young (red squares) PMS stars, together with the sample of LH 95 (\citealt{katia19}; gray filled dots).
This plot is very similar to the HR diagram (Fig. \ref{tracce}). While a separation between the old and young candidates in $T_{\rm
eff}$ is evident in the LH 95 sample  (see Fig. 9 in \citealt{katia19}), in LH 91 there is a continuous distribution in $T_{\rm eff}$, the PMS stars with the highest accretion luminosity being close to the old subgroup. 

\subsection{Mass accretion rate versus stellar age}

Finally, we derived the mass accretion rate $\dot{M}_{\rm acc}$ of our PMS candidates from the free-fall equation \citep{koenigl91,calvet98}:
\begin{equation}
L_{\rm acc} \simeq \frac{GM_\star\dot{M}_{\rm acc}}{R_\star}\left(1-\frac{R_\star}{R_{in}}\right)
,\end{equation}

\begin{figure}
   \centering
   \includegraphics[width=10cm]{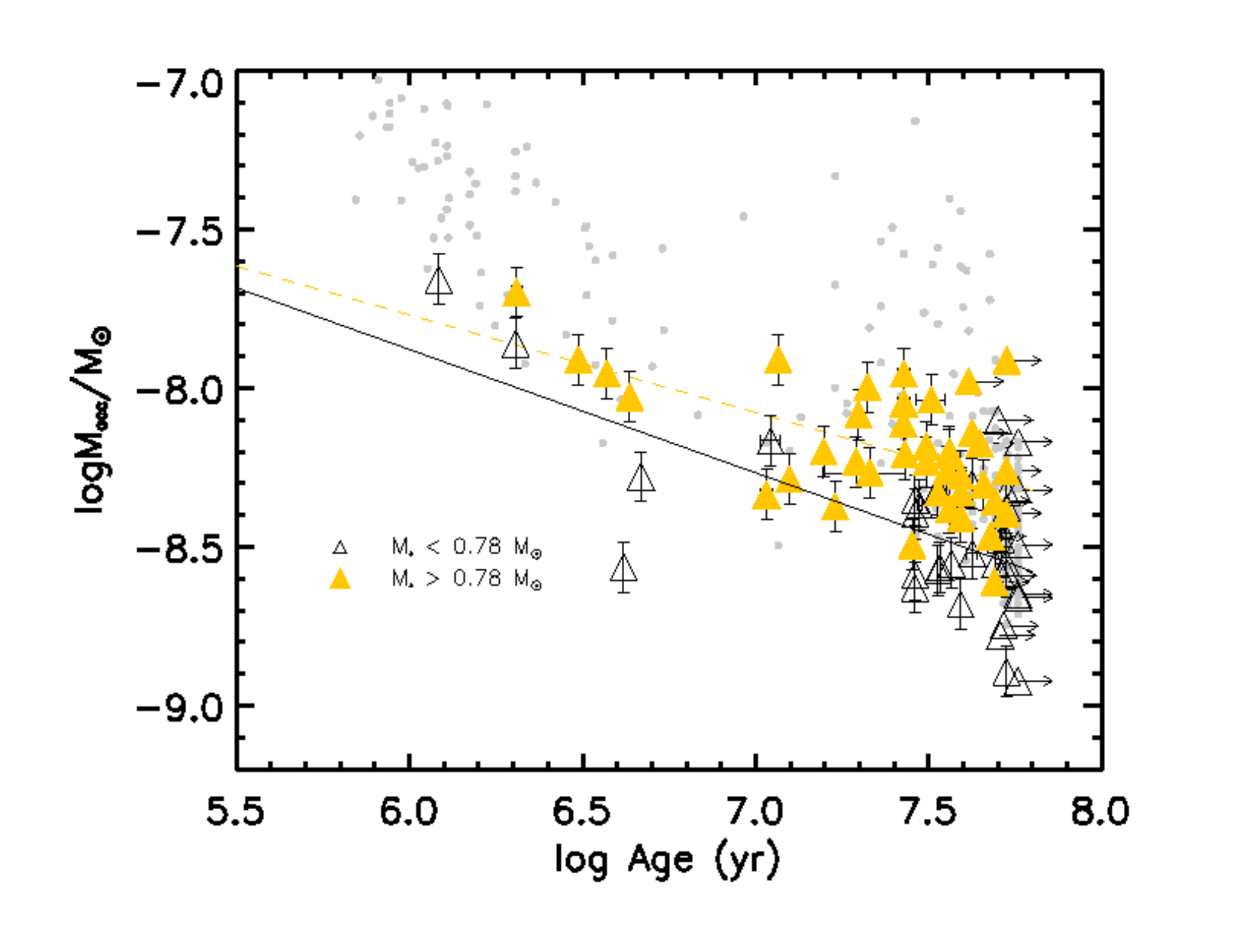}
      \caption{Observed mass accretion rate versus age  in LH 91. Yellow filled squares and  empty black triangles represent the targets with mass greater and smaller than the median mass of the PMS candidates sample, respectively. The gray-filled dots are the PMS candidates in LH 95 \citep{katia19}. The error bars on the age and mass accretion rate are reported. When the uncertainties are smaller than the symbol size, they are not visible. The arrows represent lower limits.  The  dashed yellow  and solid black lines represent the regression fit of the two subsamples of LH 91. 
      } 
   \label{tmacc}
 \end{figure}

\noindent{where $G$ is the gravitational constant, $M_\star$ and $R_\star$ are the mass and radius of the PMS candidates, and $R_{\rm in}$ is the inner radius of the accretion disk. $R_{\rm in}$ depends on how exactly the accretion disk is coupled  with the magnetic field of the star, and so its value is quite uncertain. We adopt $R_{\rm in} = 5 R_\star$, following \cite{gul98}. The median value of the mass accretion rate of our sample is $\sim$ $4.8$ $\times$ $10^{-9}$ $M_\odot$ $yr^{-1}$,  with higher values for the younger population ($\sim$ 1.2 $\times$ $10^{-8}$ $M_\odot$ $yr^{-1}$), and lower values for the older candidates ($\sim$ 4.7 $\times$ $10^{-9}$ $M_\odot$ $yr^{-1}$). The values we find are  slightly lower than those found by \cite{katia19} for LH 95, as shown in Fig. \ref{tmacc}, where the median rate  is about $7.5$ $\times$ $10^{-9}$ $M_\odot yr^{-1}$. 

The mass accretion rate in LH 91 is also lower than the median value measured in the field of SN 1987A
($2.6$ $\times$ $10^{-8}$ $M_\odot$ $yr^{-1}$, as found by \citealt{romaniello04}, and $2.9$ $\times$ $10^{-8}$ $M_\odot yr^{-1}$ as measured by \citealt{demarchi10}) and  in 30 Dor by  ($\sim$ 8 $\times$ $10^{-8}$ $M_\odot yr^{-1}$; \citealt{demarchi17}).}
The uncertainty on $\dot{M}_{\rm acc}$ is dominated by the uncertainty on $L_{\rm H\alpha}$, which is of about $16\%$, but we have to consider also the contribution of  $R_\star$ ($7\%$, including a $5\%$ systematic uncertainty on the distance modulus), stellar mass ($\sim$ $7\%$), the intrinsic uncertainties due to the evolutionary  tracks ($2\%$-$6\%$, for more details see the Appendix A of \citealt{katia19}), and knowledge of the relation $L_{\rm acc}$--$L_{\rm H\alpha}$, which in this case is not  very accurate (a factor of $\sim$ 2; \citealt{demarchi10}).
Finally, the contribution of other sources of systematic error ---such as physical processes different from accretion (e.g., chromospheric activity or ionization of nearby massive stars) or nebular continuum--- that could affect the determination of $\dot{M}_{\rm acc}$ are considered to be negligible \citep{demarchi10}. In summary, the combined statistical uncertainty on $\dot{M}_{\rm acc}$ is of about $20\%$.

A snapshot of the  mass accretion rate as a  function of the age is shown in Fig. \ref{tmacc}.
We divided the sample in two subsamples with  the stellar mass larger (yellow filled squares) and smaller  (empty black triangles) than the median stellar mass ($\sim 0.8 \,M_\odot$).
The gray filled dots are the PMS candidate in LH 95 \citep{katia19}.
As expected, the accretion appears to decrease with time, in line with the predicted evolution of viscous disks \citep{hart}, but there is a large spread  in mass accretion rate at a given age. 
We performed a linear fit to the two subsamples, and find a similar slope: $-0.31 \pm 0.07$ for the  high masses and $-0.39 \pm 0.04$ for the low masses. These values are in agreement within the error with  those evaluated in other MCs regions \citep{demarchi11,demarchi13,demarchi17,katia19}.
 This plot also shows that the mean mass accretion rate of the PMS stars in LH 91 is slightly lower than in LH 95, because our sample is composed mainly of older stars (more so than 30 Myr), close to the MS when the accretion process is less powerful.


\subsection{Mass accretion rate versus stellar mass}
\begin{figure}
   \centering
   \includegraphics[width=10cm]{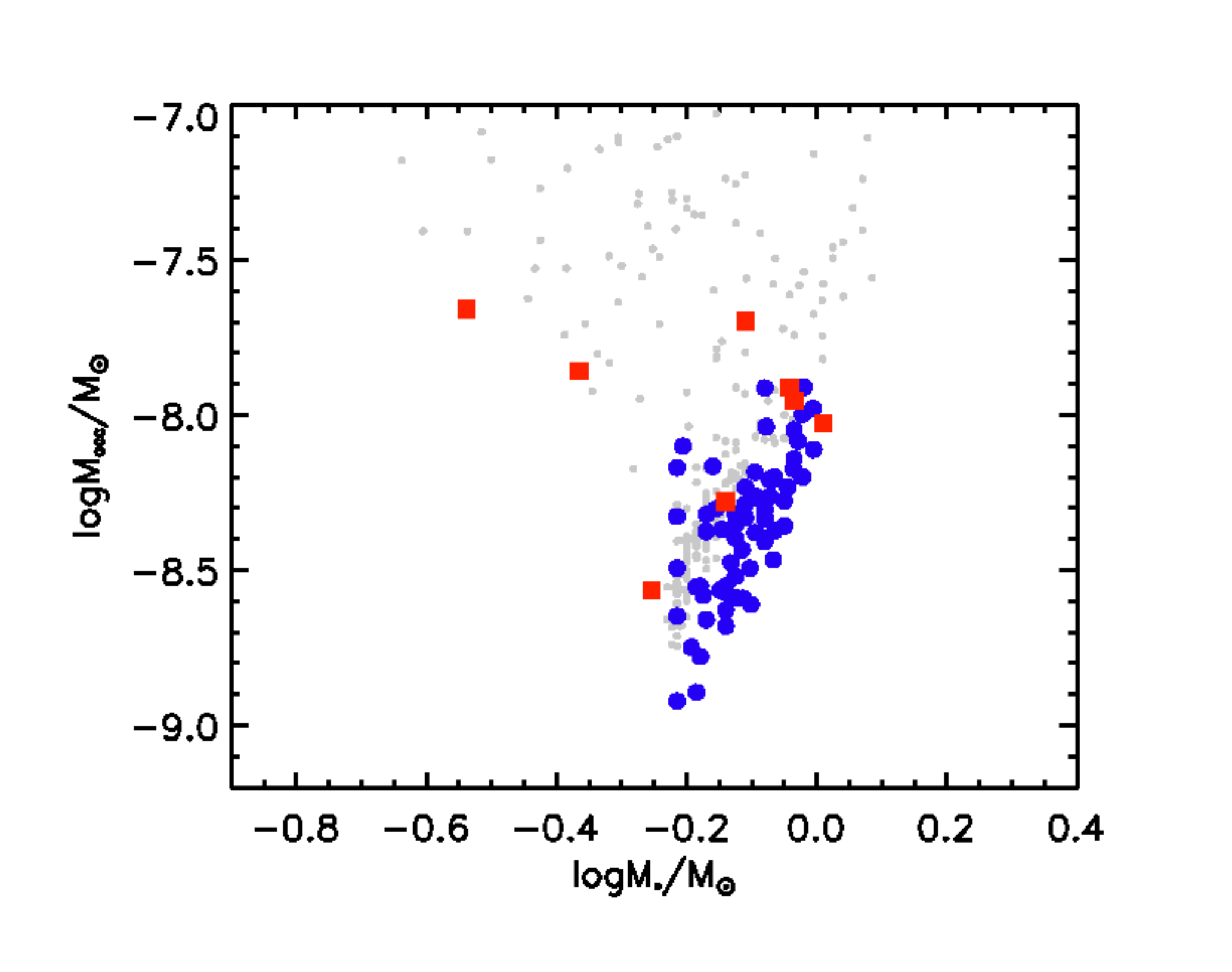}
      \caption{Distribution of $\dot{M}_{\rm acc}$ as a function of $M_{\star}$. The symbols are as in Fig. \ref{lacc}. 
}
   \label{mmacc}
 \end{figure}

Figure \ref{mmacc} shows the mass accretion rate as a function of the stellar mass of our PMS candidates. Younger PMS candidates are represented by red squares, while the older PMS candidates are marked with  blue dots.
The gray filled dots  represent the PMS candidates in LH 95 \citep{katia19}. From Figs. \ref{tmacc} and  \ref{mmacc} it is evident that the mass accretion rate is typically higher for the younger and more massive stars. Only two low-mass stars (with masses of about 0.3 $M_\odot$ and 0.4 $M_\odot$) have a high mass accretion rate (2.2 $\times$ $10^{-8}$ $M_{\odot}$/yr and 1.4 $\times$ $10^{-8}$  $M_{\odot}$/yr respectively). 
Figure \ref{mmacc} reveals a large spread in $\dot{M}_{\rm acc}$ values for a given stellar mass.
This is hardly surprising considering the large spread of ages (see also \cite{rigliaco11}).
Moreover, the older sample of PMS candidates in LH 91 reaches lower values of mass accretion rate when compared to the stars at similar masses in LH 95.
Again, it is interesting to note how the stars of any given mass appear to be younger in LH 95 than in LH 91. This could simply  be the result of different evolutionary stages between stars in LH 91 and LH 95, with the former sample being  
more evolved than the latter. This could in turn justify the smaller number of accretors and the lower values of the mass accretion rate in LH 91 compared to LH 95.
This difference might in turn be caused by other  physical differences in star formation environment, for example  the  gas density of the regions.
It would seem natural that an environment with lower gas density would result in less massive circumstellar disks, and therefore a more modest  mass accretion rate. To verify this effect, we compare the median mass accretion rate in LH 91 with that of the three star-forming regions at similar metallicity in the LMC, namely
LH 95, SN 1987A, and 30\,Dor, for which a study of accretion properties was performed. 
Considering targets with the same mass range ($0.4-1.0\,M_\odot$) and  younger than 8\,Myr, we obtained a mean $\dot{M}_{\rm acc}$ value of $1.1 \times 10^{-8}\,M_\odot$/yr for LH 91, $4.4 \times 10^{-8}\,M_\odot$/yr for LH 95, $3.7 \times 10^{-7}\,M_\odot$/yr for SN 1987A, and $5.9 \times 10^{-8}\,M_\odot$/yr for 30\,Dor.
 We also estimated the mean dust density of the aforementioned four regions taking into account the mass surface density map\footnote{https://www.asc.ohio-state.edu/astronomy/dustmaps/} by \cite{utomo}.
Considering regions with a radius of 1.5 arcmin, we found values of $0.11 \pm 0.01\,M_\odot$/pc$^2$ for LH 91, $0.16 \pm 0.01\,M_\odot$/pc$^2$  for LH 95, and similar value for SN 1987A, and 0.65 $\pm$ 0.09 $M_\odot$/pc$^2$ for 30 Dor. Even though this kind of analysis is only qualitative, we find some tentative indication that regions with higher dust densities also
have higher mass accretion rates (and possibly higher gas density), with the exception of SN 1987. A much more detailed analysis would be necessary to address this issue in more detail, which goes beyond the scope of this work.

\subsection{Spatial distribution of the PMS candidates}

\begin{figure*}
\begin{center}
\includegraphics[width=0.8\textwidth]{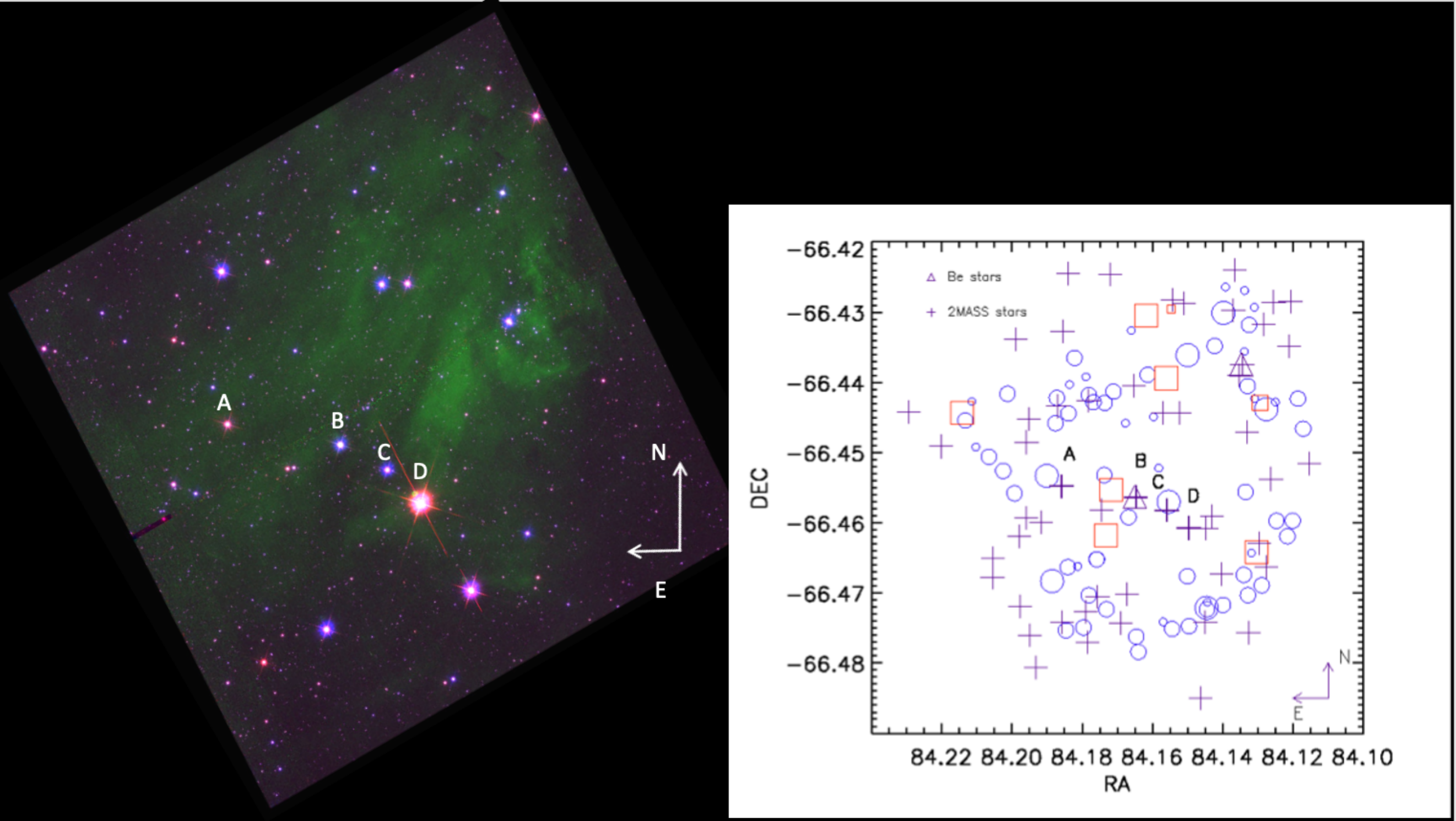}
\caption{{\it Left panel}: Color-composite image of LH 91 from WFC3 observations in the F555W, F814W, and F656N filters. The image is rotated 30 degrees to align the two figures. {\it Right panel}: Distribution of the PMS stars. Symbols are the same as in Fig. 5. The crosses represent the massive stars from  the 2MASS catalog, and the triangles are Be stars from Table\,1 of \cite{gou02}. North is up and east to the left.}
\label{space}
\end{center}
\end{figure*}

The left panel of Figure \ref{space} shows a color-composite image from WFC3 observations in the F555W (red), F814W (blue), and F656N (green) filters of LH 91. 
The right panel in the same figure shows the spatial distribution of the PMS stars in our sample projected onto the sky.
As in figure \ref{tracce}, the sizes of the dots and squares are proportionate to the ranges of the mass accretion rate. Different colors represent different ages: older than 8\,Myr (blue dots) and younger than 8\,Myr (red squares). 
The crosses are massive stars selected from the 2MASS catalog \citep{cutri} with
$J - H < 0.8$ and $J < 15$ mag (see also \citealt{katia19}), while the triangles are Be stars found by \cite{gou02} (see their Table\,1).
 To align the orientation of the two figures, the left image is rotated 30 degrees.
With the aim of  better understanding the correspondence between the fields, we indicate some bright stars with the letters A to D.
Regions with a lack of stars in the right figure  correspond to those rich in gas in the left figure, shown in green. The dust associated to the gas could be obscuring the stars behind it. The PMS objects appear to be distributed more or less uniformly over the region, and are not clustered around the massive stars, unlike the younger population of LH 95 \citep{katia19}.
This result is in agreement with the conclusions of \cite{gou02}, who found only a weak match between the HII region of LH 91 and the two Be stars located to the southwest side of the region. Therefore, it appears that in LH 91 there is no obvious region of higher star-formation intensity, at least currently.

\section{Conclusions}
\label{conclusion}
 We presented a multiwavelenght analysis of the stellar populations in LH 91, a star-forming region in the LMC, observed with the WFC3 on board the {\it HST}.
 We applied a photometric detection method to identify PMS candidates still actively accreting matter from their circumstellar disks.
 The method combines HST broad-band F555W and F814W photometry with narrow-band F656N imaging in order to identify stars with H$\alpha$ excess emission and to subsequently measure their accretion luminosity $L_{\rm acc}$ and equivalent width $EW_{\rm H\alpha}$, and to derive their mass accretion rate $\dot{M}_{\rm acc}$. 
 The main results  of our analysis can be summarized as follows:
\begin{enumerate}
\item From the photometric catalog  of 9423 well-detected stars, we identified about 180 low-mass PMS  candidates on the basis of their  excess H$\alpha$ emission,  that is, with 
their ($m_{555} - m_{656}$)  color exceeding that of the reference template at the same ($m_{555}-m_{814}$) color by more than three times the combined uncertainties on their ($m_{555}-m_{656}$)  values.
\item We measured the $EW_{\rm H\alpha}$ of the PMS stars, finding values in the range of $\sim$ 3 $\AA$ - 17 $\AA$, with a median of 9 $\AA$. We selected stars with $EW_{\rm H\alpha}$ $\geq$ 10 $\AA$, which are typical values of actively accreting PMS stars. A total of 75 objects satisfy this condition. 

\item We estimated the stellar effective temperature and luminosity thanks to the \cite{bessel} relations for $3500 \leq T_{\rm eff} \leq$ 40000 K, and the \cite{mamajek} calibrations for $T_{\rm eff} < 3500$ K.

\item We obtained the mass and age of the PMS  candidates by comparing the location of each star in the HR diagram  with theoretical PMS evolutionary tracks \citep{bressan2012}.
The range of the stellar masses in our sample is between $\sim$ 0.2 $M_\sun$ and $\sim$ 1.0 $M_\sun$ with a median of $\sim$ 0.8 $M_\sun$. The age of the stars is distributed between a few million years and as much as $\sim$ 60 Myr with an apparent gap between 5 Myr and 10 Myr.
For this reason we divided our sample in two populations, which we call younger (t $\leq$ 8 Myr with median age $\sim$ 3.5 Myr) and older PMS candidates (t $>$ 8 Myr with median age $\sim$ 35 Myr).

\item We measured the H$\alpha$ luminosity of the PMS candidates and consequently their accretion luminosity. We find a median value of $\sim$ $0.12$ $L_\sun$. The accretion luminosity increases with $L_\star$, while the dispersion in $L_{\rm acc}$ seems to decreases with $L_\star$.
We also find that the accretion luminosity spans the range 0.1-1 $L_\star$, with a peak in the distribution at about 0.3 $L_\star$. 

\item Through the accretion luminosity and other physical parameters, we determined the mass accretion rate of PMS stars, finding a median value of $\sim$ 4.8 $\times$ $10^{-9}$ $M_\sun yr^{-1}$, with higher values for the
younger population ($\sim 1.2 \times 10^{-8} M_\sun$\,yr$^{-1}$), and lower values
for the older candidates ($\sim 4.7 \times 10^{-9} M_\sun$ yr$^{-1}$).

\item We studied the relation between  the mass accretion  rate and both age and  stellar mass.  As expected, the mass accretion rate appears to decrease with time and to increase with stellar mass. 

\item We compared our results with other star formation regions in the Large Magellanic Cloud, in particular with LH 95, which is the closest region to LH 91 for which accretion properties of PMS candidates have been derived. 
LH 91 is a star-forming region that is less rich in PMS stars than LH 95, with lower stellar masses (0.2-1.0 $M_\odot$ $vs$ 0.2-1.8 $M_\odot$)
but similar range in age (few Myr up to 60Myr).
The  accretion luminosity and the mass accretion rate of PMS candidates in LH 91 are both slightly lower  than in LH 95; in particular the median values are 0.12 $L_\odot$ versus 0.17 $L_\odot$, and  7.5 $\times$ $10^{-9}$ $M_\sun yr^{-1}$ versus 4.8 $\times$ $10^{-9}$ $M_\sun yr^{-1}$ , respectively.


\item We explored the possibility that the density of the environment (which we probe using dust emission) could affect the mass accretion rate. We compared the median mass accretion rate  of star-forming regions with similar metallicity but different dust density, namely LH 91, LH 95, SN 1987A, and 30\,Dor.
 We considered targets in the same mass range (0.4-1.1 $M_\odot$) and younger than 8 Myr. From a qualitative analysis, we find that the mass accretion rate increases with dust density of the environment in which the stars are formed.
 
\item Finally, we find the spatial distribution of the PMS stars to be rather uniform, without any evidence of clumps around more massive stars.

\end{enumerate}

 The advent of the {\it James Webb Space Telescope} will allow us to put strong constraints on accretion phenomena of members in star-forming regions with different stellar properties (such as metallicity, age, and distance). In particular, the spectroscopic observations would give us information on the density and ionization state of the material undergoing accretion as well as on its kinematics, thereby providing a clearer picture of the accretion
process itself in different environmental conditions.

\begin{acknowledgements}
We are very thankful to the anonymous referee for precious comments and suggestions that have helped
us to improve this paper.
RC is grateful to ESA for the support during the data analysis useful for the preparation of this paper. This work was based on observations made with the NASA/ESA Hubble Space Telescope, and obtained from the Hubble Legacy Archive, which is a collaboration between the Space Telescope Science Institute (STScI/NASA), the Spacte Telescope European Coordinating Facility (ST-ECF/ESA) and the Canadian Astronomy Data Centre (CADC/NRC/CSA). Some of the data presented in this paper were obtained from the Mikulski Archive for Space Telescopes (MAST). STScI is operated by the Association of Universities for Research in 
Astronomy, Inc., under NASA contract NAS5-26555. This research made use of the SIMBAD database, operated at the CDS (Strasbourg, France) and data 
products from the Two Micron All Sky Survey, which is a joint project of the University of Massachusetts and the Infrared Processing and Analysis Center/California Institute of Technology, funded by the National Aeronautics and Space Administration and the National Science Foundation. This research has made also use of the SVO Filter Profile Service supported from the Spanish MINECO through grant AyA2014-55216.

\end{acknowledgements}

\bibliographystyle{aa}
\bibliography{manuscriptbib}

\end{document}